\documentclass[pas,manuscript,showpacs,showkeys,superscriptaddress,nofootinbib]{revtex4-1}
\usepackage{graphicx}
\usepackage{epsfig}  
\usepackage{epsf}    
\usepackage[tbtags]{amsmath}
\usepackage{amsfonts}
\usepackage{float}
\usepackage{subfig}
\usepackage{dcolumn}% Align table columns on decimal point
\usepackage{bm}% bold math
\usepackage{dcolumn}% Align table columns on decimal point
\usepackage{textcomp}% Align table columns on decimal point
\usepackage{multirow}
\usepackage{slashed}
\usepackage{float}
\restylefloat{figure}
\usepackage[]{hyperref}
  \hypersetup{
  bookmarks=true,         
  unicode=true,         
  pdftoolbar=true,     
  pdfmenubar=true,     
  pdffitwindow=true,    
  pdfstartview={FitH},    
  pdfsubject={Sterilenus@DUNE},  
  pdfnewwindow=true,     
  pdfcreator={RevTeX},
  colorlinks=true,     
  linkcolor=red,   
  citecolor=blue,    
  filecolor=black,  
  urlcolor=blue,           
  }
\usepackage{hypcap}

\def\nue{{\nu_e}}
\def\anue{{\bar{\nu}_e}}
\def\numu{{\nu_{\mu}}}
\def\anumu{{\bar{\nu}_{\mu}}}

\newcommand{\beq}{\begin{equation}}
\newcommand{\eeq}{\end{equation}}
\newcommand{\beqa}{\begin{eqnarray}}
\newcommand{\eeqa}{\end{eqnarray}}

\begin{document}
\title{ Non-Unitarity at DUNE and T2HK with Charged and Neutral Current Measurements}
%====================================================================================
\author{Debajyoti Dutta}
\email[Email Address: ]{debajyoti.dutta@dbuniversity.ac.in}
\affiliation{Assam Don Bosco University, Tapesia Campus, Sonapur, Assam, 782402, India}

\author{Samiran Roy}
\email[Email Address: ]{samiranroy@hri.res.in}
\affiliation{Harish-Chandra Research Institute, Chhatnag Road, Jhunsi, Allahabad 211019, India}
\affiliation{Homi Bhabha National Institute, Training School Complex,
Anushaktinagar, Mumbai - 400094, India}
%====================================================================================
%\date{\today}
%\preprint{Preprint no xxyy}

\begin{abstract}
  {Neutral current (NC) measurements play an important role in exploring  new physics scenarios at long-baseline neutrino oscillation experiments. We find that combining NC measurements of the proposed Deep Underground Neutrino Experiment (DUNE) with its charged current (CC) measurements enhances the bounds on some of the Non-Unitarity (NU) parameters. Combining DUNE with the T2HK experiment improves the bounds further. We  show that even in the averaged out regime of light sterile neutrinos, the NC events are different from the heavy sterile case in the leading order. It is observed that NC measurements at DUNE provide much better constraints on the $\alpha_{33}$ parameter than the CC measurements.  } 
\end{abstract}

%====================================================================================

\maketitle

%====================================================================================

%====================================================================================
\section{Introduction}
\label{sec1}
%==========================
The three flavor neutrino oscillation framework consists of three mixing angles $\theta_{12}$, $\theta_{13}$ and $\theta_{23}$, two mass-squared differences $\Delta m_{31}^2$ and $\Delta m_{21}^2$, and the leptonic Dirac CP phase $\delta_{cp}$. Although most of the oscillation parameters have been determined to varying degrees of precision through a combination of accelerator, solar, reactor and atmospheric neutrino experiments,  the leptonic CP phase $\delta_{cp}$ is still one of the least known parameters. Two long-baseline experiments, NO$\nu$A \citep{sanchez_mayly_2018_1286758, NOvA:2018gge} and T2K \cite{Abe:2017uxa} have been taking data in both $\nu$ and $\bar{\nu}$ modes. Although both  experiments have measured nearly the same value of $\Delta_{32}$, they have disagreements in their measurements of $\sin^2{\theta_{23}}$, hierarchy as well as $\delta_{cp}$. T2K prefers near maximal values of $\sin^2{\theta_{23}}$ in the higher octant (HO) while the  NO$\nu$A measurement, although in the HO, is  significantly higher \cite{Abe:2017uxa, Nizam:2018got}. Although both the experiments prefer normal hierarchy (NH) over inverted hierarchy (IH), yet NO$\nu$A allows IH at 1$\sigma$. The tension between the data of the two experiments is significant in an era when precision plays an increasingly important role in pointing to new physics and  disentangling it from the standard three neutrino paradigm.

Neutrino oscillation remains one of the strongest hints of physics beyond the standard model. The smallness of neutrino masses is yet to be understood and there are many different models attempting to explain it. NU of the neutrino mixing matrix \cite{ Goswami:2008mi, Antusch:2009pm,  Rodejohann:2009cq, Xing:2012kh, 
Qian:2013ora, Li:2015oal, FernandezMartinez:2007ms, Escrihuela:2015wra, Parke:2015goa, PhysRevD.34.1642, Fernandez-Martinez:2016lgt, Blennow:2016jkn, Miranda:2016wdr, Ge:2016xya, Dutta:2016vcc, Dutta:2016czj, Escrihuela:2016ube, Dutta:2016eks, Fong:2016yyh, Tang:2017khg, Soumya:2018nkw, Martinez-Soler:2018lcy, Dev:2009aw, Fong:2017gke} is yet another interesting departure from the standard three-neutrino paradigm. One way  it appears in the theory is via the type-I seesaw mechanism \cite{Mohapatra:1979ia, GellMann:1980vs, Schechter:1980gr, Valle:2015pba} which gives masses to the neutrinos via the exchange of fermionic  messengers. In the type-I seesaw, due to the presence of a Majorana mass term for heavy right handed neutrinos, we need a matrix bigger than the 3$\times$3 PMNS mixing matrix for diagonalization of the
mass matrix. This results NU of the PMNS matrix and it is discussed in detail in \cite{Fernandez-Martinez:2016lgt}. There are other variants of the low-scale seesaw mechanism such as the inverse and linear seesaw \cite{PhysRevD.34.1642,GONZALEZGARCIA1989360,Akhmedov:1995vm,Malinsky:2005bi} where the masses of the right handed neutrinos are not so heavy compared to the type-I seesaw and it can produce sizeable NU in the leptonic mixing matrix \cite{Forero:2011pc,Escrihuela:2015wra}. Generally, NU of the leptonic mixing matrix may arise due to the effect of new physics at both high and  low energy scales. At high energies, the NU of the leptonic mixing matrix, called the indirect non-unitary effect \cite{Xing:2012kh, Li:2015oal, Fong:2016yyh, Blennow:2016jkn}, is because of the mixing of heavy neutrinos with the active flavors. Such neutral heavy leptons are much heavier than the standard neutrinos. The production mechanism is the same for the two (electroweak interactions); however, if the heavy neutrinos are too massive then they cannot be produced because they are kinematically forbidden. On the other hand, direct non-unitary effects are manifested at lower energy scales \textit{i.e.} an energy which is much below the electroweak breaking scale and the mass eigenstates of light sterile neutrinos are kinematically accessible in the decays of standard model particles (such as $\pi^{+}, \pi^{-}, K^{+}$ etc.). They 
% mix with the standard neutrinos. They
also take part in neutrino oscillations.
% and hence it is possible to measure their signature in the neutrino experiments. 
It was the LSND experiment \cite{Aguilar:2001ty} which for the first time claimed a signal consistent with the oscillations driven by light sterile neutrino of $\Delta m^2_{41} \sim 1$ eV$^2$. It found an  excess of positrons  
which could be explained in terms of $\bar\nu_\mu \to \bar\nu_e$ oscillations driven by the above mass squared differences. This claim was also tested by the MiniBooNE \cite{Aguilar-Arevalo:2018gpe} experiment which ran both in the neutrino and antineutrino mode. If nature has such  sterile states, it can also lead to NU of the $3\times 3$ leptonic mixing matrix. The NU framework introduces new cp phases, which being degenerate with the standard CP phase, hamper measurements at the far detector(s) of long-baseline experiments \cite{Goswami:2008mi, Antusch:2009pm, Escrihuela:2015wra, Dutta:2016vcc, Dutta:2016czj, Escrihuela:2016ube, Dutta:2016eks, Tang:2017khg, Abe:2017jit}.

Most studies on NU have focussed on the CC measurements at the far detector of long-baseline neutrino experiments which generally measure $\nu_{\mu} \rightarrow \nu_e$ and  $\nu_{\mu} \rightarrow \nu_{\mu}$ oscillations, both in neutrino and anti-neutrino mode \cite{Escrihuela:2015wra, Dutta:2016vcc, Dutta:2016czj, Escrihuela:2016ube, Dutta:2016eks, Tang:2017khg}. Recently, neutral current measurements have been explored at DUNE \cite{Acciarri:2015uup, Acciarri:2016ooe} in the context of one light sterile neutrino \cite{Coloma:2017ptb, Gandhi:2017vzo}. Constraints on one light sterile neutrino have already been derived at DUNE and T2HK \cite{Abe:2015zbg} and can be found 
in \cite{ Berryman:2015nua, Choubey:2016fpi, Kelly:2017kch, Choubey:2017ppj}. However, the larger NU framework is more general than the one encompassing just one extra light sterile neutrino. 
%In this work,
%we mainly consider light sterile analysis to constrain NU parameters which also provie the complementary bounds for heavy sterile scenario. 

In this work, we have incorporated NC measurements with CC measurements to derive constraints on NU parameters. Already there exist tight constraints on the NU parameters \cite{Antusch:2014woa, Fernandez-Martinez:2016lgt, Escrihuela:2015wra} that come from weak interaction universality and lepton flavor violating processes (LFV). There are also model independent direct bounds on NU parameters coming from a zero distance experiment such as NOMAD \cite{Astier:2001ck, Astier:2003gs} and neutrino oscillation experiments \cite{Parke:2015goa,Qian:2013ora}. In this work, we derive the complementary bounds on NU parameters in presence of NC measurements at DUNE. We also explore the effect of combining NC measurements with CC measurements at DUNE to probe the bounds. In addition to that, we quantify the bounds that come from T2HK and then combine it with DUNE to explore the enhanced effect.

The paper is organized as follows: In section II, we review the NU framework both in the presence of heavy and light sterile neutrino, and show its effect on CC and NC measurements. In section III, we specify the details of the experiments considered in this work. In section IV, we present our results. Conclusions are drawn in section V.

\section{The NU Framework}
\label{sec3}
%==========================
In the presence of NU due to heavy sterile neutrinos, states in the mass basis ($|\nu_i\rangle$) remain orthogonal to each other, while the low energy effective flavor states \footnote{For light sterile neutrinos, the flavor basis remains orthogonal ( \textit{i.e.} $\langle \nu_{\alpha}|\nu_{\beta}\rangle =\delta_{\alpha \beta}$) as all the mass eigenstates are kinematically accessible.}($|\nu_{\alpha}\rangle$), are not orthogonal, can be represented as 
\footnote{Here we have not considered the flavor state normalization factor ($\dfrac{1}{\sqrt{(NN^\dagger)\alpha \alpha}}$) since it cancels in the event calculations \cite{Antusch:2006vwa}. \label{norm-flavor}} 
\begin{eqnarray}
|\nu_{\alpha}\rangle = N^{*}_{\alpha i}|\nu_i\rangle  ,
\label{flvor-mass}
\end{eqnarray}
where $N$ is a $3\times 3$ general matrix \cite{Escrihuela:2015wra} and can be represented as 
\begin{eqnarray}
 N = N^{ NU}  U = \begin{bmatrix}
    \alpha_{11} & 0 &0 \\
    \alpha_{21} & \alpha_{22} & 0\\
    \alpha_{31} & \alpha_{32} & \alpha_{33}
  \end{bmatrix} U.\nonumber
\end{eqnarray}
Here $U$ is the standard unitary PMNS mixing matrix and it depends on three mixing angles ($\theta_{12}$, $\theta_{13}$, and $\theta_{23}$) and one CP violating phase ($\delta_{cp}$). $N^{NU}$ contains the NU part. Under the condition that all the diagonal elements of $N^{ NU}$ are unity and all the off-diagonal elements vanish, then $N$ becomes the standard PMNS mixing matrix. The diagonal elements ($\alpha_{11}, \alpha_{22}$ and  $\alpha_{33}$) of $N^{ NU}$ are real and the off-diagonal elements ($\alpha_{21}, \alpha_{31}$ and  $\alpha_{32}$) are complex in general and can be expressed as $\alpha_{ij}=|\alpha_{ij}|e^{\phi_{ij}}$ for $i\neq j$. There are  three new CP phases $\phi_{21}, \phi_{31}$ and $\phi_{32}$ that arise in the mixing matrix $N$ in presence of NU. The new phases, especially $\phi_{21}$ can play an important role in long-baseline experiments such as DUNE and T2HK. This affects the standard CP ($\delta_{cp}$) sensitivity of these experiments significantly \cite{Dutta:2016vcc, Escrihuela:2016ube}. For the later portion of the discussion, we denote $|\alpha_{ij}|$ as $\alpha_{ij}$ for notational simplicity and mention the CP phases ($\phi_{ij}$) explicitly. In this section, we analyze the effect of NU on the neutrino oscillation probability.\\
In the presence of NU, the time evolution of the mass eigenstate in vacuum is 
\begin{eqnarray}
i\dfrac{d |\nu_i \rangle}{dt} =  H |\nu_i \rangle,
\end{eqnarray}
where $H$ is the free Hamiltonian in the mass basis and can be expressed as 
\begin{eqnarray}
 H = \begin{bmatrix}
    0 & 0 &0 \\
    0 & \dfrac{\Delta m^2_{21}}{2E} & 0\\
    0 & 0 & \dfrac{\Delta m^2_{31}}{2E} 
  \end{bmatrix}. \nonumber
\end{eqnarray}
Here $E$ is the energy of the neutrinos and $\Delta m^2_{21}$ and $\Delta m^2_{31}$ are the solar and atmospheric mass squared differences respectively. After time t($\equiv$L), the flavor state can be written as 
\begin{eqnarray}
|\nu_{\alpha}(t)\rangle =  N^{*}_{\alpha i} |\nu_i (t) \rangle =  N^{*}_{\alpha i}(e^{-iHt})_{ij}|\nu_j (t=0)\rangle.
\end{eqnarray}
Hence the transition probability from one flavor to another in the presence of NU can be written as
\begin{eqnarray}
\label{NU_prob_equn}
P(\nu_{\alpha} \rightarrow \nu_{\beta}) = |\langle \nu_{\beta}|\nu_{\alpha}(t) \rangle|^2=|N^{\ast}_{\alpha i}\, {\rm diag}(e^{-i\Delta m^{2}_{i1}t/2E})_{ij} \, N_{\beta j}|^2.
\end{eqnarray}
%%%%%%%%%%%%%%%%%%%%%%%%%%%%%%%%%%%%%%%%%%%%%%%%%%%%%%%%%%%%%%%%%%%%%%%%%%%%%%%%
Now, the transition probability for $P_{\mu e}$ with NU becomes \cite{Escrihuela:2015wra}
\begin{equation}
\label{Pme_V}
P_{\mu e} = (\alpha_{11}\alpha_{22})^2 P^{3\times 3}_{\mu e}+\alpha_{11}^2\alpha_{22} \alpha_{21}P^{I}_{\mu e}+\alpha_{11}^2 \alpha_{21}^2,
\end{equation} where $P^{3\times 3}_{\mu e}$ is the standard oscillation probability and $P^{I}_{\mu e}$ is the oscillation probability containing the new extra phase due to the NU in the leptonic mixing matrix. Here, $P^{3\times 3}_{\mu e}$ is given by
 \begin{equation}
 \begin{split}
 P^{3\times 3}_{\mu e} = 4[\cos^2\theta_{12}\:\cos^2\theta_{23}\:\sin^2\theta_{12}\:\sin^2(\frac{\bigtriangleup m^2_{21}L}{4E})+\cos^2\theta_{13}\:\sin^2\theta_{13}\:\sin^2\theta_{23}\:\sin^2(\frac{\bigtriangleup m^2_{31}L}{4E})] \\ +\sin(2\theta_{12})\:\sin\theta_{13}\:\sin(2\theta_{23})\:\sin(\frac{\bigtriangleup m^2_{21}L}{2E})\:\sin(\frac{\bigtriangleup m^2_{31}L}{4E})\:\cos(\frac{\bigtriangleup m^2_{32}L}{4E}+\delta_{cp})
 \end{split}
  \end{equation}
 
 and
  \begin{equation}
 \begin{split}
 P^{I}_{\mu e} = -2[\sin(2\theta_{13})\:\sin\theta_{23}\:\sin(\frac{\bigtriangleup m^2_{31}L}{4E})\:\sin(\frac{\bigtriangleup m^2_{31}L}{4E}+\phi_{21}+\delta_{cp})]\\
 -\cos\theta_{13}\:\cos\theta_{23}\:\sin(2\theta_{12})\:\sin(\frac{\bigtriangleup m^2_{21}L}{2E})\sin({\phi_{21}})
  \end{split}
  \end{equation}

Now from Eq. \ref{Pme_V}, we can infer:
\begin{itemize}
\item At short distances ($\dfrac{\Delta m^2 L}{E} << 1$), we will get non-zero transitions for  $\nu_{\mu} \rightarrow \nu_{e}$ in presence of NU if $\alpha_{21}$ is not zero. At such distances, the appearance probability does not depend on energy or length, thus the excess of $\nu_e$ events exactly follows the $\nu_{\mu}$ flux pattern in case of a  heavy sterile neutrino. But that is not the case for light sterile neutrino in general.
\item If $\alpha_{21} \sim 0$, we will not get any excess of $\nu_e$ events at short-baseline experiments. But even if $\alpha_{21}$ is very small, then at the far detector, the appearance probability will depend on $\alpha_{11}$ and $\alpha_{22}$. Therefore, the far detector at  long baselines will give us the unique capability to probe NU parameters as compared to the short-baseline experiments. 
\end{itemize}  

%%%%%%%%%%%%%%%%%%%%%%%%%%%%%%%%%%%%%%%%%%%%%%%%%%%%%%%%%%%%%%%
In the presence of matter, flavor eigenstates interact with matter coherently and the free Hamiltonian gets modified. The interaction Lagrangian in the presence of NU becomes:
\begin{small}
\begin{eqnarray}
\label{L_int}
\mathcal{L}_{int} = - \dfrac{g}{2\sqrt{2}} (W_{\mu} \bar{l}_{\alpha} \gamma^{\mu} (1-\gamma_5) N_{\alpha i} \nu_i) - \dfrac{g}{2 \cos(\theta_W)}(Z_{\mu}\bar{\nu}_i \gamma^{\mu} (1-\gamma_5)( N^{\dagger}  N)_{ij} \nu_j) + h.c.
\end{eqnarray}
\end{small} 
Therefore in the mass basis, the total Hamiltonian ($H_{mat}$) of the propagating neutrino is given by
\begin{eqnarray}
 H_{mat} = \begin{bmatrix}
    0 & 0 &0 \\
    0 & \dfrac{\Delta m^2_{21}}{2E} & 0\\
    0 & 0 & \dfrac{\Delta m^2_{31}}{2E} 
  \end{bmatrix} + N^{T}\begin{bmatrix}
     V_{CC} + V_{NC} & 0 &0 \\
    0 & V_{NC} & 0\\
    0 & 0 & V_{NC}
  \end{bmatrix}N^{\ast},
\end{eqnarray}
where $V_{CC}=\sqrt{2} G_F  n_e $ and $V_{NC}=-\dfrac{1}{\sqrt{2}}G_F n_n$ are the charged current and neutral current matter potential respectively. Here $n_e$ and $n_n$ are the electron and neutron densities respectively \footnote{We assume that the electron ($n_e$) and neutron ($n_n$) densities are same for DUNE and T2HK and for simplicity also assume constant matter density ($\rho=2.95$ gm/cc) for our simulated results.}. The Hamiltonian $ H_{mat}$ is hermitian and we can diagonalize it by a unitary matrix ($U_m$) as:
\begin{eqnarray}
\label{H_mat_dia}
 H_{ mat}= U_m \begin{bmatrix}
     a_1 & 0 &0 \\
    0 & a_2 & 0\\
    0 & 0 & a_3
  \end{bmatrix} U^{\dagger}_m ,
\end{eqnarray}
where $a_1 , a_2 $ and $a_3$ are the eigenvalues of $ H_{mat}$. Therefore, the transition probability ($\nu_{\alpha} \rightarrow \nu_{\beta}$) becomes
\begin{eqnarray}
P(\nu_{\alpha} \rightarrow \nu_{\beta})=|\langle \nu_{\beta}|\nu_{\alpha}(t)\rangle|^2 = | N^{\ast}_{\alpha i} ( U_{m} \, {\rm diag} (e^{-i a_1 t} , e^{-i a_2 t},e^{-i a_3 t})\,  U^{\dagger}_m)_{ij} N_{\beta j} |^2.
\end{eqnarray}
We note that most of the upcoming superbeam neutrino experiments will measure their flux through near detector (ND) measurements. Therefore, in the presence of NU the expected events at the ND differ from the actual events by a factor of $P(\nu_{\alpha} \rightarrow \nu_{\alpha})=((NN^{\dagger})_{\alpha \alpha})^{2} $. For DUNE and T2HK, the beam comprises mainly of  muon neutrinos, and hence the above factor, \textit{i.e.} the normalization factor, becomes $((NN^{\dagger})_{\mu \mu})^{2} = ((\alpha_{22})^2 + |\alpha_{21}|^2)^2$. Depending on the values of $\alpha_{22}$ and $|\alpha_{21}|$, we will get different muon events compared to (simulated) events without NU at the ND. \textit{If $\alpha_{22}\sim 0.95$ and $\alpha_{21} \sim 0$, then there will be a mismatch of around $20\%$ between the simulated and the actual events.} Thus, the ND measurements can in principle put a  tight constraint on the NU parameter $\alpha_{22}$. 

Now, the $\nu_{\alpha}$ events ($R_{\alpha}$) in a detector which are located at a distance $L$ from the $\nu_{\mu}$ source, are given by 
\begin{eqnarray}
\label{Evnts}
R_{\alpha} \sim \int dE \dfrac{d\Phi^{\rm{source}}_{\mu}(E)}{dE} \dfrac{P_{\mu \alpha} (E,L)}{L^2} \sigma_{\alpha}(E) \eta(E),
\end{eqnarray}
where $ \dfrac{d\Phi^{\rm{source}}_{\mu}(E)}{dE}$ is the muon neutrino flux at the source, $\sigma_{\alpha}(E)$ is the detection cross section and $\eta(E)$ is the detection efficiency. The flux at source is unknown. We can measure it through ND measurements, or we can rely on Monte Carlo simulation of the flux. At ND the muon events ($R_{\mu}(E)$) from $\nu_\mu$ beam can be expressed as:
\begin{eqnarray}
\label{Evnts3}
R_{\mu} (E)_{{\rm near}} \sim  dE \dfrac{d\Phi^{\rm{source}}_{\mu}(E)}{dE} (\dfrac{P_{\mu \mu} (E,L)}{L^2})_{{\rm near}} \sigma_{\mu}(E) \eta(E). \nonumber
\end{eqnarray}
Hence,
\begin{eqnarray}
 \dfrac{d\Phi^{\rm{source}}_{\mu}(E)}{dE} \propto R_{\mu} (E)_{{\rm near}} / (P_{\mu \mu})_{{\rm near}}. \nonumber
\end{eqnarray}
In the standard case $(P_{\mu \mu})_{{\rm near}} \simeq 1$ but for the NU  case $(P_{\mu \mu})_{{\rm near}} = ((NN^{\dagger})_{\mu \mu})^2 = (\alpha^2_{22} + \alpha_{21}^2)^2$ which is different from unity. So, in the presence of NU, measured flux at ND will be different from the standard simulated flux by a factor $(P_{\mu \mu})_{{\rm near}}$.
Using the ND flux measurements, the events at far detector can be expressed as:
\begin{eqnarray}
\label{Evnts1}
R_{\alpha} (E)_{{\rm far}} &\sim&  dE \dfrac{d\Phi^{\rm{source}}_{\mu}(E)}{dE} (\dfrac{P_{\mu \alpha} (E,L)}{L^2})_{{\rm far}} \sigma_{\alpha}(E) \eta(E), \nonumber \\
& \propto &  \dfrac{R_{\mu} (E)_{{\rm near}} }{ (P_{\mu \mu})_{{\rm near}}} (P_{\mu \alpha})_{{\rm far}}.
\end{eqnarray}

Therefore, we can measure the transition probability by the ratio of far and ND events as:
\begin{eqnarray}
\label{pro_ratio}
P_{\mu \alpha} \sim \dfrac{(R_{\alpha})_{{\rm far}}}{(R_{\mu})_{{\rm near}}} \sim \dfrac{(P_{\mu \alpha})_{{\rm far}}}{(P_{\mu \mu})_{{\rm near}}}.
\end{eqnarray}
For the standard case, $(P_{\mu \mu})_{{\rm near}}\simeq 1$. Hence, the measured oscillation probability ($P_{\mu \alpha}$) coincides with the actual oscillation probability $(P_{\mu \alpha})_{{\rm far}}$. But for the NU case, the measured oscillation probability ($P_{\mu \alpha}$) will be different from $(P_{\mu \alpha})_{{\rm far}}$ due to the factor $(P_{\mu \mu})_{{\rm near}}$. 
Therefore, in presence of NU, the effect of flux measurements at the ND can be included in the measured oscillation probability through the normalization factor \footnote{Note that there is no correlation between the flavor state normalization factor [\ref{norm-flavor}] and the normalization factor in the measured oscillation probability as in Eq. \ref{pro_ratio}.} $(P_{\mu \mu})_{{\rm near}}$ \cite{Blennow:2016jkn} as in Eq. \ref{pro_ratio} and that factor plays an important role in constraining the NU parameters. In the disappearance channel, the NU parameter which causes the maximum effect in the probability cancels the effect of that NU parameter when we consider the normalization factor as discussed in \cite{Hernandez-Garcia:2017pwx}. It is also apparent from the disappearance plot of Fig. \ref{fig1}.

Now, if we use the simulated flux at the source, then using Eq. \ref{Evnts}, we can directly calculate the transition probability.  Hence, for the simulated flux, we do not need to consider the normalization factor $((NN^{\dagger})_{\mu \mu})^2$ (arsing from the ND flux measurements) to measure the transition probability and the measured oscillation probability ($P_{\mu \alpha}$) will be same as $(P_{\mu \alpha})_{{\rm far}}$. In this analysis, we have considered two cases \textit{i.e.} one with the simulated flux at source where normalization factor is not required in the probability expression and the other with ND measurements \textit{i.e.} the normalization factor is present in the probability calculation.  

\begin{figure}[!htb]
\includegraphics[width=0.35\textwidth]{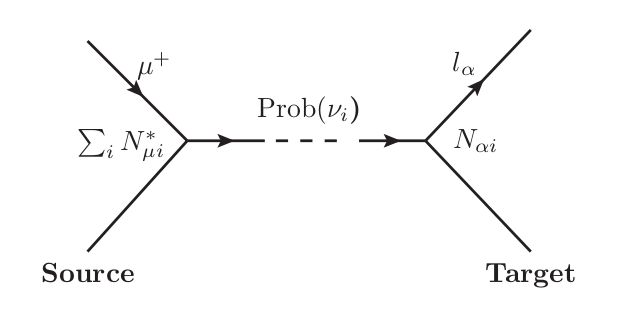} 
\includegraphics[width=0.35\textwidth]{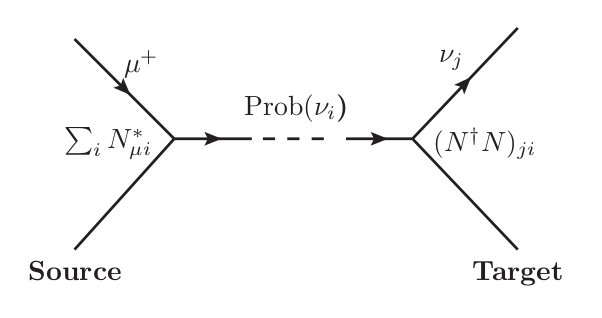}
\caption{ Schematic representation of CC and NC events for $\nu_{\mu}$ beam in the presence of NU. Here Prob($\nu_i$) = $exp(-i \, m^{2}_i L/2E)$, where L is the distance between source and target and E is the energy of the neutrino.}
\label{Sch_CC_NC}
\end{figure}

In the presence of NU, both CC and NC events get modified. The schematic representation of CC and NC events are shown in Fig. \ref{Sch_CC_NC} and can be written as

\begin{eqnarray}
N_{events}^{CC} & \sim & \int dE \frac{d\phi^{cc, SM}_{\mu}(E)}{dE} |\sum^{3}_{i=1}A(W \rightarrow \mu^{+} \nu_{i}) exp(-i \Delta m^2_{i1}L/2E) A(\nu_{i} W \rightarrow l_{\alpha})|^2 \sigma^{CC}_{\alpha} \eta \nonumber \\
&=&  \int dE \frac{d\phi^{cc, SM}_{\mu}(E)}{dE} |\sum^{3}_{i=1} N^{*}_{\mu i} exp(-i \Delta m^2_{i1}L/2E) N_{\alpha i}|^2 \sigma^{CC}_{\alpha} \eta \nonumber \\
&=& \int dE \frac{d\phi^{cc, SM}_{\mu}(E)}{dE} P(\nu_{\mu} \rightarrow \nu_{\alpha}) \sigma^{CC}_{\alpha} \eta 
\end{eqnarray} 
where $A$ stands for amplitude, $\sigma_{\alpha}^{CC}$ is the CC cross section for $\nu_{\alpha}$, and $\eta$ is the efficiency of the detector. The NC events for $\nu_{\mu}$ beam in vacuum are 

\begin{eqnarray}\nonumber
N_{events}^{NC} &\sim & \int dE \frac{d\phi^{cc, SM}_{\mu}(E)}{dE} \sum^{3}_{j=1}|\sum^{3}_{i=1}A(W \rightarrow \mu^{+} \nu_{i}) exp(-i \Delta m^2_{i1}L/2E) A(\nu_{i} Z \rightarrow \nu_j)|^2 \sigma^{NC} \eta \\ 
& = & \int dE \frac{d\phi^{cc, SM}_{\mu}(E)}{dE}  \sum^{3}_{j=1}|\sum^{3}_{i=1}N^{*}_{\mu i} exp(-i \Delta m^2_{i1}L/2E) (N^{\dagger}N)_{ji}|^2 \sigma^{NC} \eta,
\end{eqnarray}
where  $\sigma^{NC}$ is the NC cross section. Hence the NC events can be written as
\begin{eqnarray}
\label{NC_Heavy}
N^{NC}_{events} &\propto & \sum^{3}_{j=1}|\sum^{3}_{i=1}N^{*}_{\mu i} exp(-i \Delta m^2_{i1}L/2E) (N^{\dagger}N)_{ji}|^2.
\end{eqnarray}
In the presence of matter, NC events will be proportional to
\begin{eqnarray}
N^{NC}_{events}  \propto  \sum^{3}_{k=1}|\sum^{3}_{i,j=1}N^{*}_{\mu i} ( U_{m}\, {\rm diag} (e^{-i a_1 t} , e^{-i a_2 t},e^{-i a_3 t})  U^{\dagger}_m)_{ij} \, (N^{\dagger}N)_{kj}|^2,
\end{eqnarray}
where $U_m$ and $a_i$'s are defined in Eq. \ref{H_mat_dia}.

\subsection*{Light Sterile case:}

The $3\times 3$ PMNS mixing matrix will also become non unitary in the presence of light sterile neutrinos. As shown in \cite{Blennow:2016jkn}, in this case, the leading charged current transition probability among the active flavors will remain the same as the non unitary case (due to the heavy sterile neutrino) if the effect of the light sterile neutrino is averaged out in the detector. The active flavor states in the presence of light sterile neutrinos can be represented as
\begin{eqnarray}
|\nu_{\alpha} \rangle  =  u^{*}_{\alpha l}|\nu_{l}>=\sum^{3}_{i =1} N^{*}_{\alpha i}|\nu_i \rangle + \sum^{n}_{J =4}\Theta^{*}_{\alpha J}|\nu_J \rangle,
\end{eqnarray}
where $u$ is a unitary mixing matrix and its dimension (n) depends on the number of sterile neutrinos. $N$ represents the  $3 \times 3$ active-light sub-block of $u$ and $\Theta$ represents the  $3\times n$ sub-block of $u$ that mixes active and heavy states. The vacuum transition probability $\nu_{\alpha} \rightarrow \nu_{\beta}$ in presence of light sterile neutrino is given by %\cite{Blennow:2016jkn}
\begin{eqnarray}
P(\nu_{\alpha} \rightarrow \nu_{\beta})& =&| \langle \nu_{\beta}|\nu_{\alpha}(t) \rangle|^2 =|u^{*}_{\alpha i}\, {\rm diag}(e^{-i\Delta m^{2}_{i1}t/2E})_{ij} \, u_{\beta j}|^2\nonumber \\ 
& =&|\sum^{3}_{i,j=1} N^{*}_{\alpha i} \, {\rm diag}(e^{-i\Delta m^{2}_{i1}t/2E})_{ij} \, N_{\beta j}  +  \sum^{n}_{J,K=4}\Theta^{*}_{\alpha J}\, {\rm diag}(e^{-i\Delta m^{2}_{J1}t/2E})_{JK} \, \Theta_{\beta K}|^2\nonumber \\
& = &|\sum^{3}_{i,j=1} N^{*}_{\alpha i}\, {\rm diag}(e^{-i\Delta m^{2}_{i1}t/2E})_{ij} \, N_{\beta j}|^2 + \mathcal{O}(\Theta^4).
\label{Pme_light_sterile}
\end{eqnarray}
The cross terms will vanish in the limit $\Delta m^{2}_{J1}L/2E >>1$ (where $L\equiv t$) since the finite energy resolution of the detector will render  $<\sin(\Delta m^{2}_{J1}L/2E)>=<\cos(\Delta m^{2}_{J1}L/2E)>=0$. Therefore, if we neglect the correction corresponding to order ($\Theta^4$), then the leading order transition probability $ \nu_{\alpha} \rightarrow \nu_{\beta}$, will be same as Eq. \ref{NU_prob_equn}. 

In the presence of light sterile neutrinos, the neutral current events also change from their standard value. Only the active flavors participate in the neutral current events, and we have, for the vacuum case and a $\nu_{\mu}$ beam the proportionality to
\begin{eqnarray}
N^{NC}_{events} &\propto &\sum^{n}_{j=1}|\sum^{n}_{i=1}A(W \rightarrow \mu^{+} \nu_{i}) exp(-i \Delta m^2_{i1}L/2E) A(\nu_{i} Z \rightarrow \nu_j)|^2\nonumber \\
\label{NC_light}
& = & \sum^{n}_{j=1}|\{\sum^{n}_{i=1}u^{*}_{\mu i} exp(-i \Delta m^2_{i1}L/2E) (\sum_{\rho= e, \mu , \tau}u^{\dagger}_{j \rho} u_{\rho i})\} |^2  \\
& = & \sum_{\rho= e, \mu , \tau} P(\nu_{\mu} \rightarrow \nu_{\rho}).
\end{eqnarray}
Now from Eq. \ref{NC_light}, we can write 

\begin{small}
\begin{eqnarray}
N^{NC}_{events} \propto \sum^{n}_{j=1}|\{\sum^{n}_{i=1}u^{*}_{\mu i} exp(-i \Delta m^2_{i1}L/2E) (\sum_{\rho= e, \mu , \tau}u^{\dagger}_{j \rho} u_{\rho i}) \}|^2  \nonumber
\end{eqnarray}
\begin{eqnarray}
& = &  \sum^{n}_{j=1}|\{\sum^{3}_{i=1}N^{*}_{\mu i} exp(-i \Delta m^2_{i1}L/2E) (\sum_{\rho= e, \mu , \tau}u^{\dagger}_{j \rho} N_{\rho i}) + 
 \sum^{n}_{I=4}\Theta^{*}_{\mu I} exp(-i \Delta m^2_{I1}L/2E) (\sum_{\rho= e, \mu , \tau}u^{\dagger}_{j \rho} \Theta_{\rho I})\}|^2 \nonumber \\ 
& = &  \sum^{n}_{j=1}|\{\sum^{3}_{i=1}N^{*}_{\mu i} exp(-i \Delta m^2_{i1}L/2E) (\sum_{\rho= e, \mu , \tau}u^{\dagger}_{j \rho} N_{\rho i})\}|^2 + \mathcal{O}(\Theta^4) \nonumber \\
& \simeq &  \sum^{3}_{j=1}|\{\sum^{3}_{i=1}N^{*}_{\mu i} exp(-i \Delta m^2_{i1}L/2E) (\sum_{\rho= e, \mu , \tau}N^{\dagger}_{j \rho} N_{\rho i})\}|^2 + \sum^{n}_{J=4}|\{\sum^{3}_{i=1}N^{*}_{\mu i} exp(-i \Delta m^2_{i1}L/2E) (\sum_{\rho= e, \mu , \tau}\Theta^{\dagger}_{J \rho} N_{\rho i})\}|^2 \nonumber \\
\label{NC_light1} 
& = &  \sum^{3}_{j=1}|\{\sum^{3}_{i=1}N^{*}_{\mu i} exp(-i \Delta m^2_{i1}L/2E) (N^{\dagger}N)_{ji}\}|^2 + \sum^{n}_{J=4}|\{\sum^{3}_{i=1}N^{*}_{\mu i} exp(-i \Delta m^2_{i1}L/2E) (\sum_{\rho= e, \mu , \tau}\Theta^{\dagger}_{J \rho} N_{\rho i})\}|^2. 
\end{eqnarray}
\end{small}

Due to the presence of the $\Theta^2$ term in Eq. \ref{NC_light1}, the neutral current events will not remain same as Eq. \ref{NC_Heavy} in the leading order. Therefore, the NC analysis will be different for light and heavy sterile case in the leading order. 

The constraints on NU parameters due to the heavy sterile are very tight \cite{Antusch:2014woa, Fernandez-Martinez:2016lgt}.  
The constraints derived by using precision measurements of electroweak processes are not applicable for light sterile neutrinos. However, oscillation experiments provide the best way to probe the NU in the presence of light sterile neutrinos. In the averaged out regime of light sterile neutrino corresponding to Eq. \ref{Pme_light_sterile}, the NU due to a heavy sterile is nearly same as light sterile neutrino. Therefore, we take the mass of the light sterile neutrino to be such that it averages out before reaching the ND. Henceforth, in the rest of the paper, we consider NU that comes from light sterile neutrino. Using NC and CC measurements, we constrain NU parameters that also provide the complementary bounds of NU parameters due to the heavy sterile neutrinos.

 %as the oscillation probabilities nearly same in both cases.}

%==========================
\section{Experimental and Simulation details}
\label{sec2}
%==========================
In this work, we present our calculations for the DUNE and T2HK experiments. The specifications of these experiments are as follows:
\subsection{DUNE}
%%%%%%%%%%%

 DUNE \cite{Acciarri:2015uup} is a
proposed future superbeam experiment at Fermilab, U.S.A capable of establishing the existence of CPV in the leptonic sector. It is also capable of resolving  issues like mass hierarchy and the octant of $\theta_{23}$. Its optimized beam of 1.07 MW - 80 GeV protons will deliver $1.47\times 10^{21}$
protons-on-target (POT) per year. The far detector will be placed at the Homestake mine in South Dakota. It is a Liquid Argon (LAr) detector of mass 40 Kt and the baseline is 1300 km. The experiment will run for 7 years dividing its time equally between neutrinos and anti-neutrinos, which corresponds to a total exposure of $4.12\times 10^{23}$
kt-POT-yr. All the details of the experiment like CC signal and background definitions,  detector efficiencies are taken from \cite{Alion:2016uaj}. The details of the NC events at DUNE are taken from \cite{Adams:2013qkq}.  The assumed detection efficiency related to  NC event is 90$\%$. In a NC event, the outgoing (anti-)neutrino
carries away a fraction of the incoming energy. Due to this missing energy, the
reconstructed visible energy is on the average less than the total incoming energy. Hence, using a gaussian energy resolution function cannot give accurate results. We have used  migration matrices from \cite{DeRomeri:2016qwo} to reproduce the NC event spectra correctly. We also consider 10$\%$ of the CC events as NC background in $\nu$ (and $\bar{\nu}$) mode.  For the
NC  analysis, we take  5$\%$
and 10$\%$ signal and background normalization errors respectively. Other details regarding the NC measurements are taken from \cite{Gandhi:2017vzo}.

%%%%%%%%%%%%%%
\subsection{T2HK}
%%%%%%%%%%

 Hyper-Kamiokande (HK) \cite{Abe:2015zbg, Abe:2016ero, Abe:2018uyc} is the upgraded version of the Super-Kamiokande (SK) \cite{ Fukuda:1998mi} program in Japan. In this experiment, the fiducial mass of the SK detector will be increased by about twenty times. HK will have two 187 kt third generation Water Cherenkov detector modules which will be placed near the current SK site. The detector will be placed  at a baseline of 295 km from the J-PARC proton accelerator research complex in Tokai, Japan. T2HK has almost similar physics goals as DUNE, such as measuring the neutrino mass hierarchy, the octant of $\theta_{23}$, and determining the leptonic CP phase.

In our analysis we have considered a beam power of 1.3 MW and the 2.5$^{0}$ off-axis flux for T2HK. The total fiducial mass considered is 374 kt, correponds to two tanks each of 187 kt. We have assumed a total run time of 10 years, of which the  neutrino run will be  for 2.5 years while the  anti-neutrino run will be for 7.5 years. The assumed energy resolution is $15\%/\sqrt{E}$. As a check, we have matched the number of events used in this work with the TABLE III  and TABLE IV of ref. \cite{Abe:2016ero}. The signal normalization error in $\numu(\anumu)$ disappearance and $\nue(\anue)$ appearance channel are 3.9\% (3.6\%) and 3.2\% (3.6\%) respectively. The background and energy calibration errors assumed in this work are 10\% and 5\%, respectively for all channels.

%%%%%%%%%%%%%%

Throughout the analysis, we have fixed the true values or the best fit values of the neutrino oscillation parameters as given in \cite{Esteban:2018azc} unless stated. We fix the true values of the solar and the reactor mixing angles at $\theta_{12}=33.82^{\circ}$ and $\theta_{13}=8.61^{\circ}$ respectively. The assumed true value of the atmospheric mixing angle is $\theta_{23}=49.7^{\circ}$. The true value of the leptonic CP phase is fixed at $\delta_{cp}=217^{0}$. The mass square differences considered in this work are $\Delta m^2_{21}=7.39\times 10^{-5}$ eV$^2$ and $\Delta m^2_{31}=2.525\times 10^{-3}$ eV$^2$ respectively. The 95$\%$ CL bounds on the NU parameters are taken from the neutrino experiments only and can be found at \cite{Blennow:2016jkn}.
%\footnote{Since in the averaged out limit of light sterile neutrino, the NU due to heavy sterile is same as light sterile neutrino and hence the bounds given in \cite{Blennow:2016jkn} is also applicable to our case.}.
We have prepared a NU code for this work which provides results consistent with the MonteCUBES \cite{Blennow:2009pk} NU engine. The results presented in this work are generated by incorporating our NU code with GLoBES \cite{Huber:2007ji,Huber:2004ka}. 

\section{Results}
\label{results}
In this section, we present our results for DUNE and T2HK experiments. First, we discuss the effect of NU on neutrino oscillation at the probability level and then at the $\chi^2$ level. 

\subsection{Probability Plots}
\label{pro-sec}
\begin{figure}[!htb]
\hspace{-01.5cm} \includegraphics[width=0.5\textwidth]{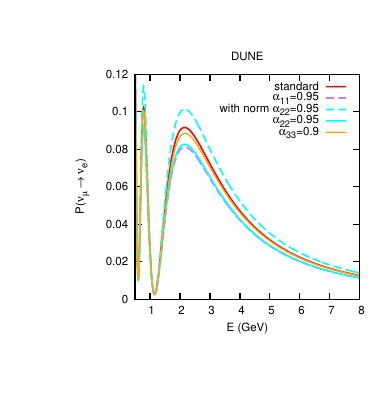} 
\includegraphics[width=0.5\textwidth]{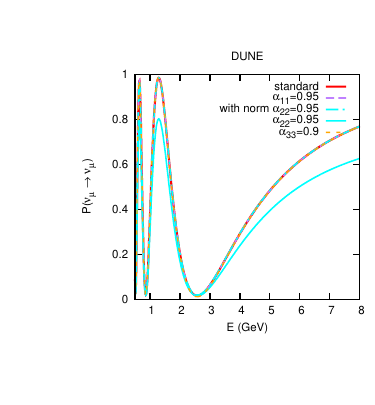}
\vspace{-01.2cm} 
\caption{ Effect of diagonal NU parameters on appearance and disappearance channels considering one parameter at a time. All the non diagonal parameters are kept at zero. The plots are shown for $\alpha_{11}=\alpha_{22}=0.95$ and $\alpha_{33}=0.9$.}
\label{fig1}
\end{figure}

\begin{figure}[htb]
\hspace{-01.5cm}\includegraphics[width=0.5\textwidth]{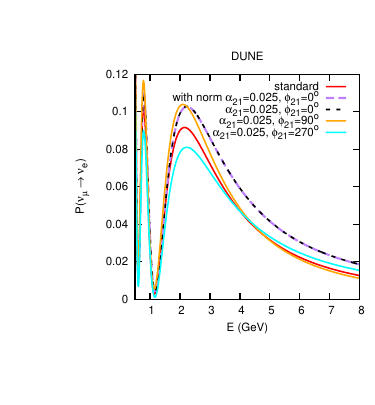}
\includegraphics[width=0.5\textwidth]{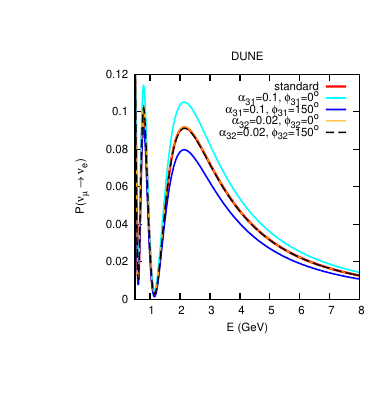}
\center
\vspace*{-3.0cm}
\includegraphics[width=0.5\textwidth]{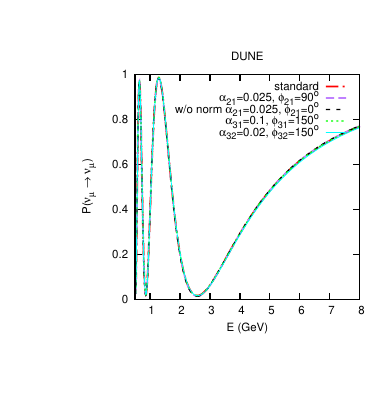}
\vspace*{-1.80cm}
\caption{Effect of non-diagonal NU parameters on the appearance and disappearance channels.}
\label{fig2}
\end{figure}

In Fig. \ref{fig1} and Fig. \ref{fig2}, we show the effect of NU parameters on both the appearance and the disappearance probabilities at DUNE  in presence of matter effects. We consider one NU parameter at a time to disentangle the effect of a particular parameter from the rest but in the $\chi^2$ analysis we consider all the parameters as simultaneously varying. Again, to incorporate the ND measurements, we have to consider the normalization factor $((NN^{\dagger})_{\mu \mu})^2 = (\alpha^2_{22} + \alpha_{21}^2)^2$ \footnote{Only $\alpha_{22}$ and $\alpha_{21}$ will arise in the normalization factor for the $\nu_{\mu}$ beam. Therefore, we consider the normalization factor for $\alpha_{22}$ and $\alpha_{21}$. We consider only one NU parameter at a time while generating the probability plots. Hence, there is no difference between with and without normalization for other NU parameters.}, in the transition probability as in Eq. \ref{pro_ratio}. But that is not the case for the simulated flux. We show the probability plots both with and without the normalization factor. Wherever we use the normalization factor, we specify it in the plots.

Fig. \ref{fig1} shows the effects of the diagonal NU parameters on the appearance and disappearance channels. The red line corresponds to the standard $3\nu$ oscillation probability. The purple line corresponds to the case with $\alpha_{11}=0.95$. It is seen that $\alpha_{11}$ has a significant effect on the appearance channel. The cyan solid (dashed) line shows the effect of $\alpha_{22}$ with (without) the normalization factor. In the appearance channel, $\alpha_{22}$ has a significant effect irrespective of the normalization factor. On the other-hand, normalization reduces the effect of $\alpha_{22}$ on the disappearance channel. But the effect of $\alpha_{22}$ without the normalization is significant in the disappearance channel. Therefore, if we consider the simulated flux, then both the appearance and disappearance channels will get affected by $\alpha_{22}$. The effect of $\alpha_{33}$ is very small on both the appearance and the disappearance channel and hence constraining it by these channels is not very fruitful.

In Fig. \ref{fig2}, we show the effect of non diagonal NU parameters on the oscillation probability while setting the diagonal parameters to unity. Since there is a phase associated  with each non-diagonal parameter, we show the probability plots for a fixed value of the phase $\phi_{ij}$.  The top left panel shows the variation of $\alpha_{21}$ while the top right panel shows the  variation of $\alpha_{31}$ and $\alpha_{32}$. In the lower panel, we have shown the effect of all the three non diagonal parameters on the disappearance channel. Even a small value of $\alpha_{21}$ can change $\rm P(\nu_{\mu} \rightarrow \nu_e)$ oscillation probability significantly almost for all the values of energy. The effect of the normalization factor is negligible in this case. If the phase $\phi_{21}$ is allowed to vary for a given value of $\alpha_{21}$, the probability deviates from the standard 3$\nu$ case specially around the oscillation maxima. The other two non-diagonal parameters $\alpha_{31}$ and $\alpha_{32}$ has negligible effect on the appearance channel. But for larger values of $\alpha_{31}$ (say around 0.1), we can see a significant deviation from the standard case and a large phase dependency.
From the plots in the lower panel, it is observed that the non diagonal parameters do not affect the measurements of the disappearance channel significantly. From all  of these results, we can draw the following conclusions:

\begin{figure}[!htb]
\hspace{-01.5cm} \includegraphics[width=0.54\textwidth]{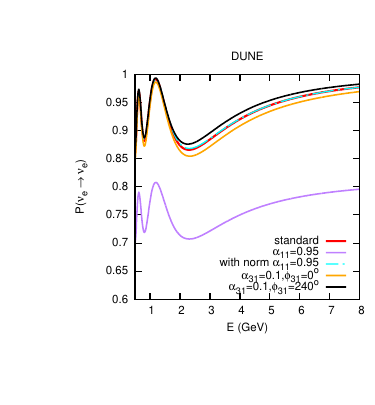} 
\hspace{-01.5cm} \includegraphics[width=0.54\textwidth]{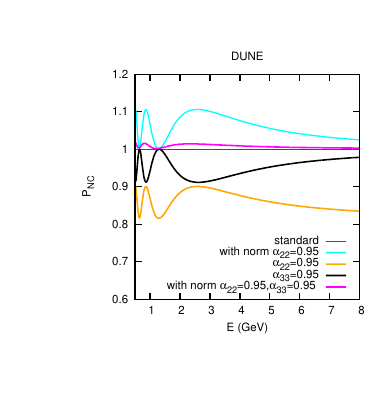}
\vspace{-01.2cm} 
\caption{ Effect of NU parameters on the Background neutrino oscillation ($\nu_e \rightarrow \nu_e$) and on the neutral current probability.}
\label{fig3}
\end{figure}

\begin{itemize}
\item Effects of $\alpha_{11}$ on the appearance channels are large compared to the disappearance channel. 

\item The effect of normalization is crucial for the $\alpha_{22}$ parameter. Depending on the normalization condition both the appearance and disappearance channel will contribute.

\item The effect of $\alpha_{33}$ is very small on both the appearance and disappearance channel.

\item Out of the three non diagonal parameters, $\alpha_{21}$ affects the appearance probability significantly. The effect is enhanced in presence of the phase.  None of these parameters has any noticeable effect on disappearance probability.
\end{itemize}
In Fig. \ref{fig3}, we show the variation of $P_{ee}$  as well as the NC measurements with energy as a function of NU parameters. The $P_{ee}$ oscillation probability plays an important role in the background events. With the normalization,\footnote{For $P_{ee}$ channel the normalization factor is $\alpha^{4}_{11}$.} although the effect of $\alpha_{11}$ on $P_{ee}$ is negligible, yet the probability changes drastically for the same value of $\alpha_{11}$ if the normalization is switched off. The effect of normalization on $P_{ee}$ channel plays an important role to constrain the $\alpha_{11}$ parameter as discussed in \ref{chisq-analysis} and Appendix \ref{margi-sec}. The effect of $\alpha_{31}$ is small compared to $\alpha_{11}$ but it shows a mild CP dependence. From the right panel of Fig. \ref{fig3}, we observe that in the presence of $\alpha_{22}$ and $\alpha_{33}$, the $P_{NC} \,\,(=P_{\mu e} + P_{\mu \mu} +P_{\mu \tau})$ oscillation probability decreases significantly from unity. But with normalization factor (as $\alpha^{4}_{22}$ is in the denominator) the NC probability becomes greater than unity. Therefore, when we consider both the parameters $\alpha_{22}$ (with norm) and $\alpha_{33}$ simultaneously there is a cancellation between $\alpha_{22}$ and $\alpha_{33}$ as shown by the pink line in the right panel of Fig. \ref{fig3}. 

\begin{figure}[!htb]
\includegraphics[width=0.42\textwidth]{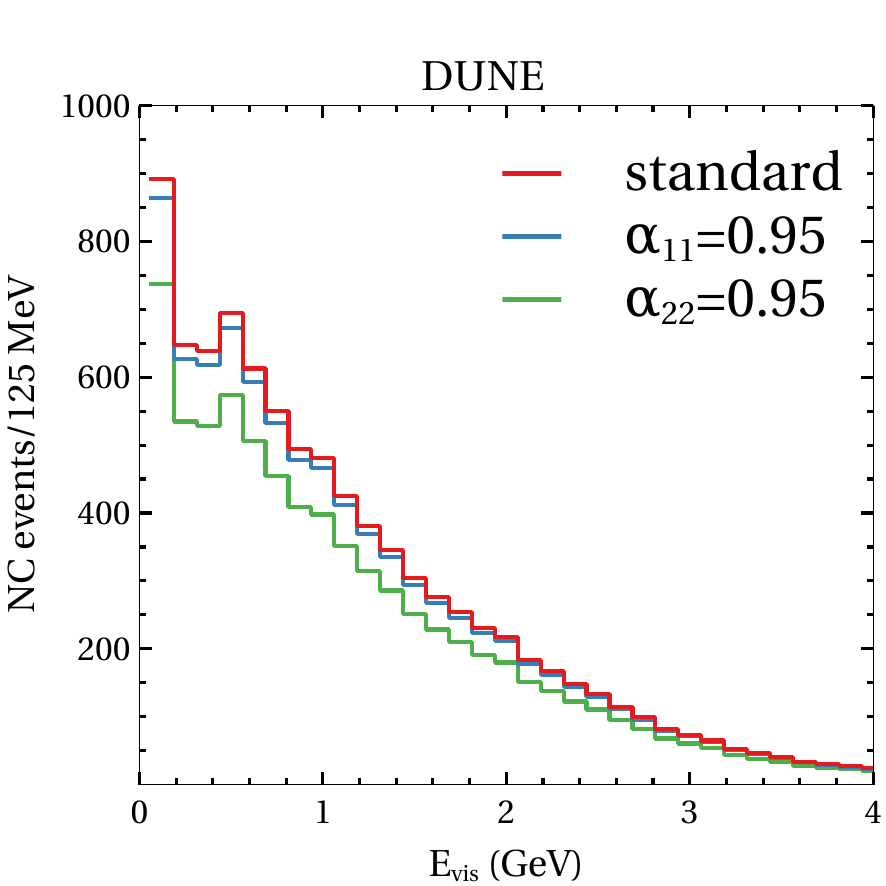} 
 \includegraphics[width=0.42\textwidth]{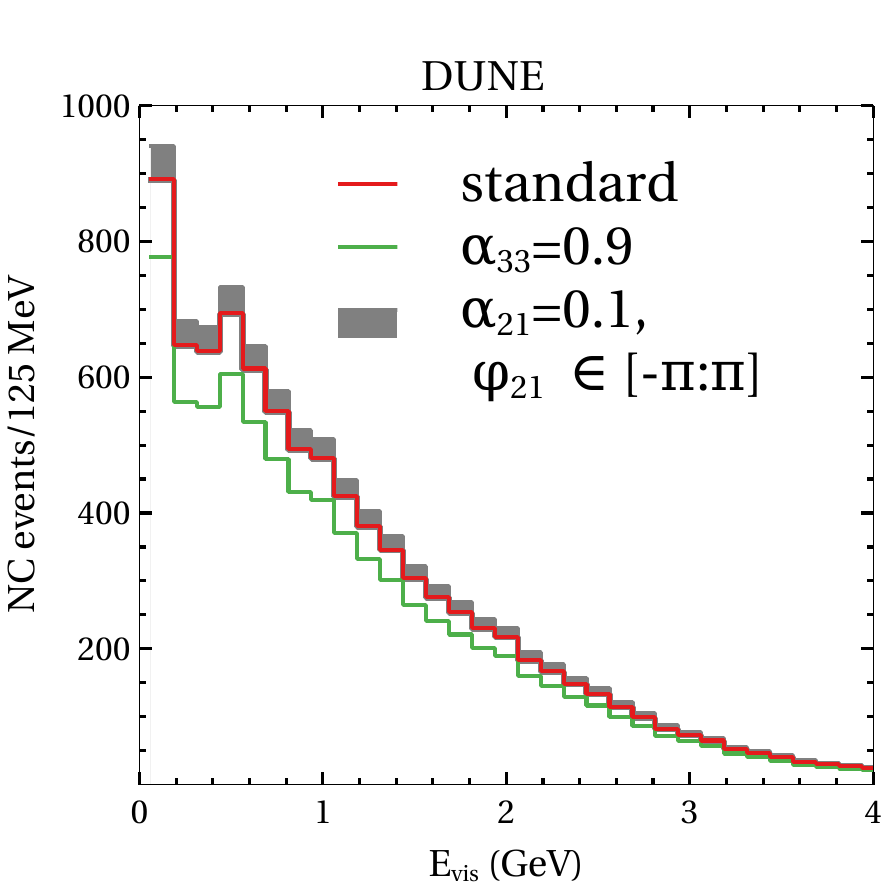} 
\caption{NC events for neutrino in the presence of one NU parameter at a time without the normalization factor.}
\label{events-nc}
\end{figure}
\subsection{NC event Plots}
\label{nc-events-sec}
In Fig. \ref{events-nc}, we show the NC events as a function of the visible energy (E$_{\rm{vis}}$) in the detector in presence of NU parameter without the normalization factor. We consider one NU parameter at a time to show the effect of that particular parameter in the NC events. The red line corresponds to the expected number of events in the standard scenario. From the left panel, we observe that in presence of $\alpha_{22}$, the NC events reduce significantly. Therefore, NC measurements are useful to constrain the $\alpha_{22}$ parameter. The effect of $\alpha_{11}$ is very mild on NC events as shown by the blue line in the left panel. In the right panel, we have shown the effect of $\alpha_{33}$, $\alpha_{21}$ and $\phi_{21}$. In presence of $\alpha_{33}$, the events get suppressed from the standard expected one. Hence, we can put an effective bound on $\alpha_{33}$ using NC measurements. The grey band corresponds to the case of varying $\phi_{21}$ in the full range keeping $\alpha_{21}$ at $0.1$. The band crosses the standard events since we consider only $\alpha_{21}$ with $\phi_{21}$ variation keeping other parameters fixed to their standard model values. Now, $\alpha_{21}$ affects the NC events mildly and there are CP dependency in the events. Therefore, NC helps to improve the bound on $\alpha_{21}$ slightly.

In the next subsection, we present our sensitivity plots to constraint the NU parameters.

\subsection{$\chi^2$ analysis}
\label{chisq-analysis}

To quantify the effect of NU at DUNE and T2HK, we have performed a $\chi^2$ analysis. 
We define the $\chi^2$ as: 

\begin{eqnarray}
\label{chi2}
\chi^2 (\textbf{n}^{{\rm true}}, \textbf{n}^{{\rm test}}, f) = 2 \sum_{i}^{N_{reco}} (n_{i}^{{\rm true}} ln \dfrac{n_{i}^{{\rm true}}}{n_{i}^{{\rm fit}} (f)} + n_{i}^{{\rm fit}} (f) - n_{i}^{{\rm true}}) + f^2,
\end{eqnarray}
where $\textbf{n}$ are event rate vectors in $N_{reco}$ bins of reconstructed energy and $f$ represents the nuisance parameter.
$n_{i}^{{\rm true}}$ stands for the true events corresponding to standard three neutrino oscillation paradigm and $n_{i}^{{\rm fit}} (f)$ represents the events corresponding to new physics \textit{i.e.} NU. 
In the fit, we have marginalized over all the standard neutrino oscillation parameters in their $1 \sigma$ CL allowed ranges.  The standard CP phase ($\delta_{cp}$) is marginalized  over the full range. In addition to that, we have also marginalized over all NU parameters in the ranges \cite{Blennow:2016jkn}: $\alpha_{11}\in[1, 0.976], \alpha_{22} \in [1, 0.978], \alpha_{33} \in [1, 0.9], \alpha_{21}\in [0, 0.025], \alpha_{31}\in [0, 0.069]$ and
$\alpha_{32}\in [0, 0.012]$. The unknown cp phases $\phi_{ij}$ are marginalized over the full range \textit{i.e.} $\phi_{ij}\in [0^{\circ}, 360^{\circ}]$. In this way, we choose the minimum $\Delta \chi^2$ for a selective NU parameter by marginalizing over all the standard as well as the remaining NU parameters.

\begin{figure}[!htb]
%\hspace{-01.5cm}
\includegraphics[width=0.45\textwidth]
{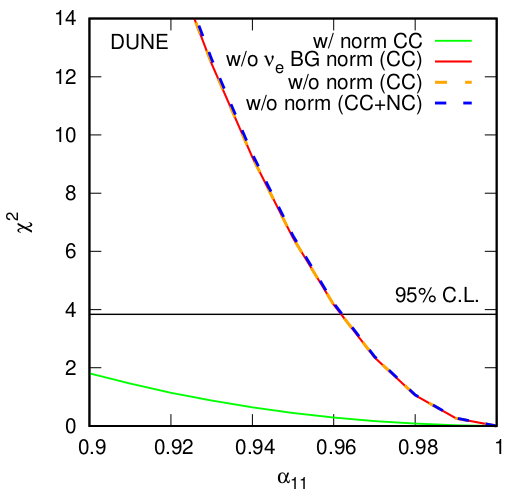} 
\includegraphics[width=0.44\textwidth]
{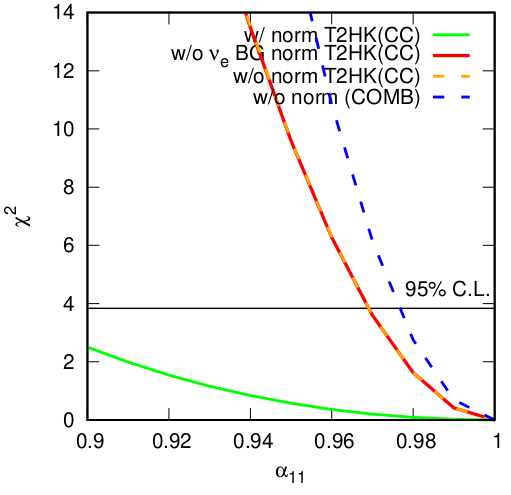}
\includegraphics[width=0.45\textwidth]
{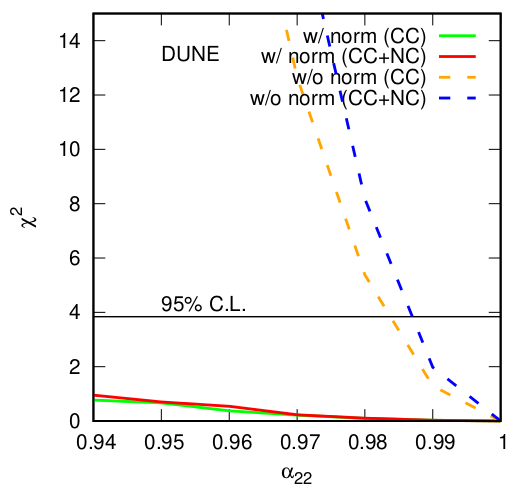}
\includegraphics[width=0.44\textwidth]
{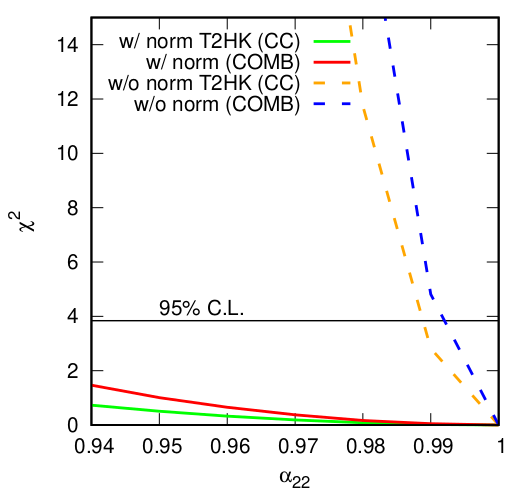}
%\vspace*{-1.5cm}
\caption{Constraints on $\alpha_{11}$ and $\alpha_{22}$ at DUNE and T2HK using CC measurements. We have also combined CC measurements at T2HK with the CC+NC measurements at DUNE. We call these combined results as `COMB'.}
\label{fig4}
\end{figure}

\begin{figure}[!h]
\hspace{-01.5cm}\includegraphics[width=0.45\textwidth]
{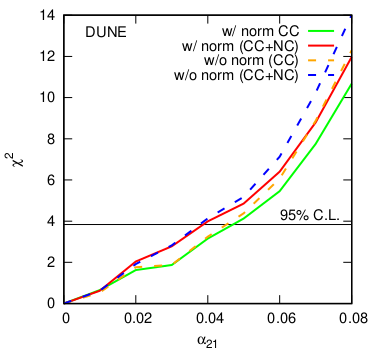} 
\includegraphics[width=0.45\textwidth]
{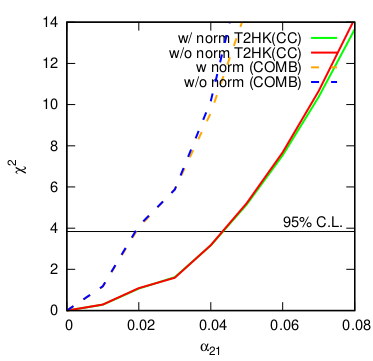}
%\vspace*{-1.5cm}
\caption{Constraints on $\alpha_{21}$ at DUNE and T2HK using the CC measurements. Also  shown are  the effects of combining CC and NC measurements at DUNE, and finally, the  combination of  DUNE (CC+NC) with the CC measurements at T2HK.}
\label{fig5}
\end{figure}

\begin{figure}[!h]
\hspace{-01.5cm}\includegraphics[width=0.45\textwidth]
{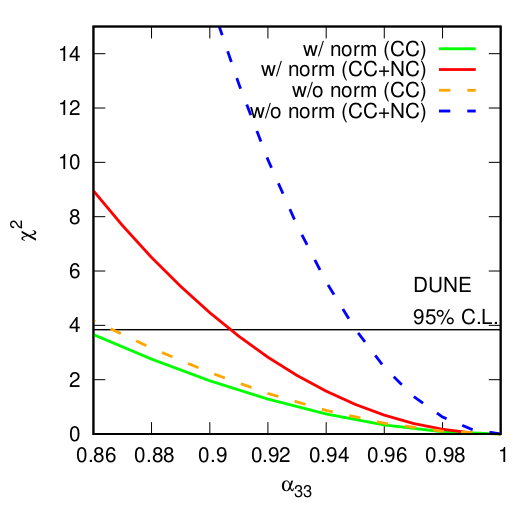} 
\includegraphics[width=0.425\textwidth]
{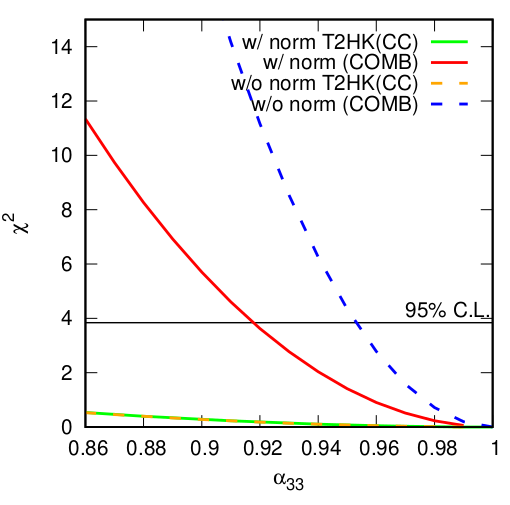}
%\vspace*{-1.5cm}
\caption{Constraints on $\alpha_{33}$ at DUNE and T2HK using CC measurements. We have also combined CC measurements at T2HK with the CC+NC measurements at DUNE.}
\label{fig6}
\end{figure}

In Fig. \ref{fig4}, \ref{fig5}, and \ref{fig6}  we show the capability of DUNE, T2HK and their combination, to probe the NU parameters. We focus mainly on the diagonal NU parameters $\alpha_{11}$, $\alpha_{22}$, and $\alpha_{33}$, and one off-diagonal parameter $\alpha_{21}$. The bounds on the other two off-diagonal parameters \textit{i.e.} $\alpha_{31}$ and $\alpha_{32}$ are not improved compared to the present bounds and hence those results are not included in this analysis. The results are shown for two specific cases: 
(I)  with normalization factor (w/ norm) \textit{i.e.} the flux is determined through ND measurements, and
(II) without normalization factor (w/o norm) \textit{i.e.} the simulated flux is used at source. 
The plots captioned as `w/ norm' means that the norm factor is used for both background and signal. The term `w/o $\nu_e$ BG norm' stands for the case where the norm factor is not used for the $\nu_{e}$ (and $\bar{\nu}_{e}$) background, but used for all other backgrounds. The term `w/o norm' stands for the cases where norm factor is not used for both background and signal. In the top panel of Fig. \ref{fig4} we show the sensitivity of $\alpha_{11}$ (upper panel) and $\alpha_{22}$ (lower panel) both for DUNE and T2HK. We present the results for CC measurements at T2HK and then for the combination of it with CC and NC measurements at DUNE, named as `COMB'. In Fig. \ref{fig5}, we also show the constraints for $\alpha_{21}$ parameter and in Fig. \ref{fig6}, obtained constraints are shown for $\alpha_{33}$ both at DUNE and T2HK. We draw the following conclusions from this analysis:\\

\paragraph{\underline{Bound on $\alpha_{11}:$}} It is observed from the upper panel of Fig. \ref{fig4} that both DUNE and T2HK give weak constraints when we use the norm factor in the measurements. But, if the norm factor is not used in the $\nu_e$ (and $\bar{\nu}_{e}$) background, then we can see a significant enhancement in the sensitivity in both DUNE and T2HK. This enhancement is because of the decrease in $\nu_{e}$ (and $\bar{\nu}_{e}$) background due to the exclusion of the norm factor. This can also be confirmed from the probability plots in Fig. \ref{fig3}. The disappearance channel $\nu_e \rightarrow \nu_e $ (or $\bar{\nu}_e \rightarrow \bar{\nu}_e $) depends significantly on $\alpha_{11}$. Hence the effect of marginalization of other parameters like $\alpha_{21}$ does not affect this channel. This point is also discussed in the Appendix \ref{margi-sec}. Thus, it is important to point out that the $\nu_e$ (and $\bar{\nu}_{e}$) background is the main channel for constraining the $\alpha_{11}$ parameter. The NC measurements of DUNE do not improve the bound further. From the plots, we observe that DUNE can exclude all values of $\alpha_{11} \leq 0.962$ while T2HK can exclude all $\alpha_{11} \leq 0.97$ at 95$\%$ CL. Thus, T2HK provides  slightly better constraint on $\alpha_{11}$ compared to DUNE. Combination of the two experiments can improve the constraint further and at 95$\%$ CL, it can exclude all $\alpha_{11} \leq 0.978$. 
%\red{We also observe that in DUNE, adding NC has no effect on $\alpha_{11}$ as NC+CC case gives same constraint on $\alpha_{11}$ as that of CC only.}

\paragraph{\underline{Bound on $\alpha_{22}:$}} From the lower panel of Fig. \ref{fig4}, we observe that like the previous case both DUNE and T2HK can not put strong constraint on $\alpha_{22}$ when measured with the norm factor. Even after combining the two experiments, the bound on $\alpha_{22}$ does not improve with the norm factor. But it improves significantly when the measurements are done without the norm factor. DUNE (T2HK) CC measurements can exclude all $\alpha_{22} \leq 0.983$ ($\alpha_{22} \leq 0.989$) at 95 $\%$ CL. As shown in Fig. \ref{events-nc}, the NC events at DUNE get reduced in the presence of $\alpha_{22}$. Therefore, when we add NC measurements with the CC measurements at DUNE, we observe an enhancement on the bound and all $\alpha_{22} \leq 0.987$ can be ruled out at 95$\%$ CL. Finally, when we combine both the experiments we get a very tight constraint on $\alpha_{22}$ and  at 95$\%$ CL, all $\alpha_{22} \leq 0.992$ can be ruled out. 
%\red{ Since the disappearance probability for $\alpha_{22}$ decreases significantly in the absence of the norm factor (see Fig. \ref{fig1}), we can conclude that the constraint on $\alpha_{22}$ mainly comes from the disappearance channel.}

\paragraph{\underline{Bound on $\alpha_{21}:$}} From Fig. \ref{fig5}, we observe that the use of the norm factor does not affect the bounds on $\alpha_{21}$, unlike $\alpha_{11}$ and $\alpha_{22}$. CC measurements at DUNE (T2HK) can constrain all $\alpha_{21}\geq 0.046$ ($\alpha_{21}\geq 0.042$) at 95$\%$ CL. Adding NC with CC at DUNE improves the constraint further. Combining T2HK with DUNE raises the constraint significantly and all $\alpha_{21}\geq 0.019$ can be excluded at 95$\%$ CL\footnote{An improved bound on $\alpha_{21}$ parameter can be achieved in short-baseline experiments at Fermilab and related details can be found in \cite{Miranda:2018yym}.}.

\paragraph{\underline{Bound on $\alpha_{33}:$}} From Fig. \ref{fig6}, we note that NC measurements at DUNE can improve the bounds on $\alpha_{33}$ compared to CC measurements. In the presence of $\alpha_{33}$, NC events decrease significantly compared to the standard expected values as shown in Fig. \ref{events-nc}. Therefore, combining CC measurements with NC measurements at DUNE improves the bound on $\alpha_{33}$ to a great extent. Without the norm factor, combining NC with CC measurements at DUNE constrains $\alpha_{33}$ such that at 95$\%$ CL all values of $\alpha_{33} \leq 0.95$ are excluded. Use of the norm factor alleviates the sensitivity as the marginalization over $\alpha_{22}$ cancels the effect of $\alpha_{33}$ as shown by the pink line in the right panel of Fig. \ref{fig3}. For T2HK, CC measurements do not improve the bounds on $\alpha_{33}$. DUNE CC measurements give better bounds on $\alpha_{33}$ compared to T2HK both with and without the norm factor due to the large matter effect. Combination of this with CC+NC measurements at DUNE slightly improves the bounds. The bound on $\alpha_{33}$ that comes from the combination is $\alpha_{33}\leq 0.952$ at 95$\%$ CL.

\subsection{Effect of higher systematic}
\begin{figure}[htb!]
\includegraphics[width=0.45\textwidth]
{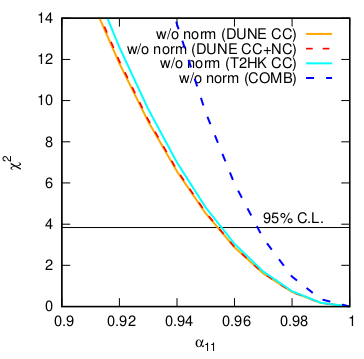} 
\includegraphics[width=0.46\textwidth]
{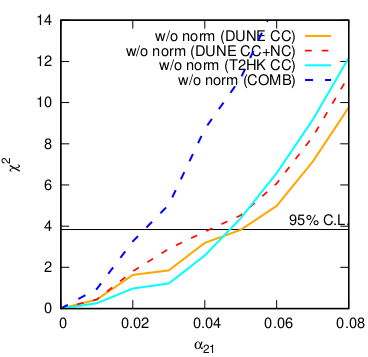}
\includegraphics[width=0.46\textwidth]
{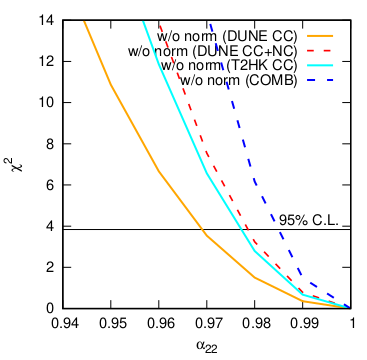}
\includegraphics[width=0.45\textwidth]
{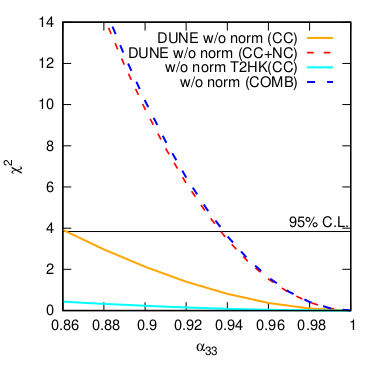}
\caption{Constraints on $\alpha_{11}$, $\alpha_{21}$ $\alpha_{22}$ and $\alpha_{33}$ at DUNE and T2HK with higher systematic uncertainties. We have also combined CC measurements at T2HK with the CC+NC measurements at DUNE. We call these combined results as `COMB'.}
\label{higher_syst}
\end{figure}
%  and show them by the blue dotted line. The orange, blue and cyan dashed plots are for DUNE (CC), DUNE (CC+NC) and T2HK (CC) respectively.

As we know that the simulated flux by Monte Carlo is typically `tuned' to experimental data and hence in the absence of the ND measurements, the systematic uncertainties will be larger than the assumed systematic in the section \ref{sec2}. The $\chi^{2}$ plots presented in the previous section are generated with the standard values of systematic uncertainties associated with DUNE and T2HK experiments even in the absence of ND flux measurements. Here in this section, we have shown the constraints on $\alpha_{11}$, $\alpha_{21}$, $\alpha_{22}$ and $\alpha_{33}$ with higher values of the systematic uncertainties in the absence of ND \textit{i.e.} the scenario where we do not consider the norm factor in the probability expression. We have considered a signal and a background error of 10$\%$ for both DUNE and T2HK experiments and hence re-calculated the sensitivities for the following cases: DUNE (CC), DUNE (CC+NC), T2HK (CC) and DUNE (CC+NC)+T2HK (CC). We observe from Fig. \ref{higher_syst} that the sensitivity to a particular parameter deteriorates due to the increase in systematic uncertainties. 
With these increased systematic uncertainties, the $95 \%$ CL exclusion bounds on NU parameters coming from the combination of DUNE and T2HK (`COMB' case) are $\alpha_{11} <0.968, \, \, \, \alpha_{22} <0.985, \, \, \alpha_{33} < 0.936, $ and $\alpha_{21}>0.023$. 

\begin{table}[h!]
 \begin{tabular}{|c|c|c |c|c|} 
 \hline
 NU parameters & Present Bounds  & W/O Norm & W/ Norm  & W/O Norm \\  
  & at 95$\%$ CL & (Normal systematic) & (Normal systematic) & (Large systematic) \\ 
 \hline
 $\alpha_{11}$ & 0.976 & 0.978 & - & 0.968 \\ 
 \hline
 $\alpha_{22}$ & 0.978 & 0.992 & - & 0.985 \\
 \hline
 $\alpha_{33}$ & 0.9 & 0.952 & 0.917 & 0.936 \\
 \hline
 $\alpha_{21}$ & 0.025 & 0.019 & 0.019 & 0.023 \\ 
 \hline
\end{tabular}
\caption{Present bounds \cite{Blennow:2016jkn} on the NU parameters and the bounds coming from the combination of DUNE and T2HK for different scenarios.}
\label{tab}
\end{table}

The constraint on NU parameters derived in \cite{Blennow:2016jkn} rules out all $\alpha_{11}<0.976$, $\alpha_{22}<0.978$, $\alpha_{33}<0.9$, and $\alpha_{21}>0.025$ at 95$\%$ CL. However, the combination of DUNE and T2HK can improve the NU bounds further. The combination of DUNE and T2HK without the norm factor can rule out  all $\alpha_{11}<0.978$, $\alpha_{22}<0.992$, $\alpha_{33}<0.952$, and $\alpha_{21}>0.019$ at 95$\%$ CL. But with the norm factor, the ability of DUNE and T2HK to constrain $\alpha_{11}$ and $\alpha_{22}$ is reduced significantly and we are not able to improve the constraints on these parameters. But with the norm factor, we can still improve the constraints on $\alpha_{21}$ and $\alpha_{33}$. The combination of DUNE and T2HK can rule out   all $\alpha_{21}>0.019$, and $\alpha_{33}<0.917$ at 95$\%$ CL. With the increase of systematic uncertainties, the capability of the experiments to constrain the NU parameters reduces to an extent as shown in the Table \ref{tab}.

\section{Summary and concluding remarks}

In this work, we have studied the constraints on NU parameters at DUNE and T2HK, especially focusing on the contributions of NC measurements at DUNE. We consider the NU that comes from a light sterile neutrino as the NU due to a heavy sterile is severely constraint from the electroweak precision measurements. But these constraints are not applicable for a light sterile neutrino and the oscillation experiments give us the best way to probe it. In the averaged out regime of light sterile neutrino, the NU due to a light sterile is equal to that of a heavy sterile neutrino for CC measurements in the leading order. In contrast, NC events in the presence of both heavy and light sterile neutrino differ from each other in the leading order.

In this analysis, we have considered two scenarios of the flux measurements \textit{i.e.} one with the ND measurements (w/ norm case) and the other one with the simulated flux at source (w/o norm case). With the simulated flux, the combine analysis of DUNE and T2HK improves the constraint on NU parameters. But with the ND measurements, the capability to constrain NU parameters reduces significantly. However, in this case, the combination of DUNE and T2HK can improve the bounds on $\alpha_{21}$ and $\alpha_{33}$ from present values. We also show our results with higher systematic uncertainties in the case of simulated flux at source. Even with 10$\%$ signal and background uncertainties, the combination of DUNE and T2HK provides better bounds on NU parameters except $\alpha_{11}$.
We have found that the $\nu_{e}$ background oscillation is the most dominant channel in the measurements of $\alpha_{11}$ parameter and hence this parameter will be better constrained by this background than the signal itself. In the case of $\alpha_{22}$, NC measurements help in improving the bounds further. Finally, we have found that NC measurements at DUNE help in deriving better bounds on $\alpha_{33}$ parameter compared to the CC measurements.

\section*{Acknowledgements} 
We acknowledge Boris Kayser and Raj Gandhi for their valuable suggestions. We are grateful to Michel Sorel for providing us with the migration matrices used in this
work. The work of SR is supported by INFOSYS scholarship for senior students. We acknowledge the use of HRI cluster facility to carry out the computations in this work.

%\section{Appendices}
\appendix
\setcounter{equation}{0}
\setcounter{section}{0}
\renewcommand{\thesection}{\Alph{section}}
\renewcommand{\theequation}{\thesection\arabic{equation}}
\section{Effect of Marginalization and Normalization Factor} 
%-----------------------------
\label{margi-sec}
%-----------------------------
\begin{figure}[h]
\hspace{-01.5cm}\includegraphics[width=0.45\textwidth]
{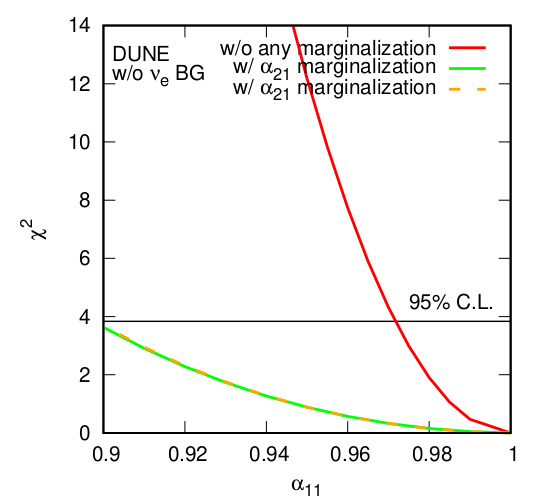} 
\includegraphics[width=0.43\textwidth]
{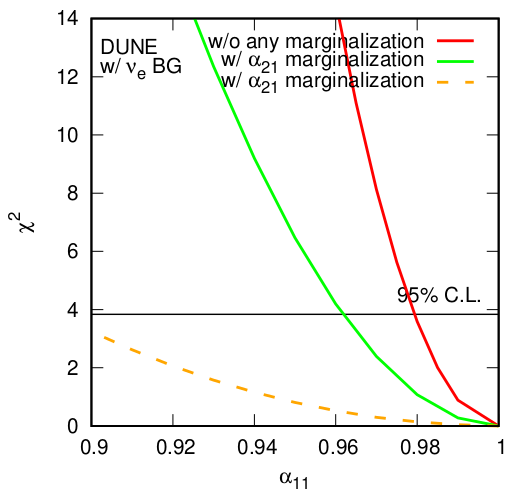}
%\vspace*{-1.5cm}
\caption{Constraint on $\alpha_{11}$ using the appearance channel.  The solid line corresponds to the case without normalization factor and the doted line corresponds to the case with normalization factor.}
\label{Pme_marg}
\end{figure}

In Fig. \ref{Pme_marg}, we have shown the ability of the appearance channels to constrain $\alpha_{11}$ parameter and also discuss the effect of marginalization of $\alpha_{21}$ parameter. In the left panel, we consider the appearance channels without any $\nu_e$ ($\bar{\nu}_{e}$) background in the simulation but in the right panel, we show the effect of $\nu_e$ ($\bar{\nu}_{e}$) background in the $\chi^2$ analysis. The solid line corresponds to the case without any normalization factor but the dotted line corresponds to the case with the normalization factor. We observe from the left panel that when we do not consider any marginalization then the appearance channels put tight constraint on $\alpha_{11}$. But if we consider the marginalization over both $\alpha_{21}$ and $\phi_{21}$, then the ability of the appearance channels to constrain $\alpha_{11}$ reduces drastically. In this case, the effect of normalization factor is not significant as shown by the green solid line and yellow dotted line. In the right panel, we have considered the $\nu_e$ ($\bar{\nu}_{e}$) background in the simulation. Now, the disappearance channels ($\nu_e \rightarrow \nu_e$ or $\bar{\nu}_e \rightarrow \bar{\nu}_e$) do not depend on other parameters significantly except $\alpha_{11}$. Therefore, when we perform the marginalization over $\alpha_{21}$ and $\phi_{21}$, the contribution that comes from $\nu_e$  or $\bar{\nu}_{e}$ background is still significant as shown by the green solid line. But the use of the normalization factor cancels the effect of $\alpha_{11}$ on $\nu_e \rightarrow \nu_e$ (and $\bar{\nu}_e \rightarrow \bar{\nu}_e$) oscillation probability. Hence, with normalization it is not possible to put tight bounds on $\alpha_{11}$ as shown by the yellow doted line. The same argument holds for $\alpha_{22}$ where the main contribution comes from the disappearance channels $\nu_{\mu} \rightarrow \nu_{\mu}$ and $\bar{\nu}_{\mu} \rightarrow \bar{\nu}_{\mu}$ respectively.

%In the left panel, the $\nu_e$ background contribution is not considered but in the right panel the contribution from $\nu_e$ background is included.

\bibliographystyle{apsrev}
\bibliography{NU-bib}
\end{document}